\documentclass[aip,apl,reprint,floatfix,nobalancelastpage,english]{revtex4-1}
\usepackage{graphicx}
\usepackage[dvips]{color}
\usepackage{color} 
\usepackage{babel}
\usepackage{amsmath} 
\usepackage{hyperref} 
\usepackage{amssymb}

\begin{document}

\title{Protecting superconducting qubits from external sources of loss and heat}

\author{Antonio D. C\'orcoles}
\author{Jerry M. Chow} 
\author{Jay M. Gambetta} 
\author{Chad Rigetti} 
\author{J. R. Rozen} 
\author{George A. Keefe} 
\author{Mary Beth Rothwell} 
\author{Mark B. Ketchen}
\author{M. Steffen} 

\affiliation{IBM T.J. Watson Research Center, Yorktown Heights, New York 10598, USA}

\date{August 5, 2011}

\begin{abstract}
We characterize a superconducting qubit before and after embedding it along with its package in an absorptive medium. We observe a drastic improvement in the effective qubit temperature and over a tenfold improvement in the relaxation time up to 5.7~$\mu$s. Our results suggest the presence of external radiation inside the cryogenic apparatus can be a limiting factor for both qubit initialization and coherence. We infer from simple calculations that relaxation is not limited by thermal photons in the sample prior to embedding, but by dissipation arising from quasiparticle generation.

\end{abstract}


\maketitle 

Energy loss in superconducting qubits remains a major object of study on the road towards scalable qubit architectures. The primary origins of relaxation in superconducting qubits can be intrinsic to the material, encompassing dielectric,\cite{Martinis2005} and resistive (quasiparticle) losses,\cite{Catelani2011,Lenander2011} or external through radiative\cite{Sage2011,WennerAPS} and electromagnetic (EM) environmental losses.\cite{schoelkopf02,Houck2008} The current understanding of these loss mechanisms is still incomplete, as it is difficult to experimentally separate the different contributions to qubit relaxation. It is equally critical to consider the bath to which the qubits relax, as the equilibrium temperature of such a bath determines the degree of purity for qubit initialization.\cite{walls_quantum_1995,nielsen_chuang_2000} Thermal heating of qubits is well known in experiments,\cite{YalePvtComm} although not often pointed out, and its elimination is necessary for future quantum computing applications. 

In this Letter, we present an experiment where extrinsic loss mechanisms on a superconducting qubit are removed, resulting in both a lower effective qubit temperature and a significant increase in qubit coherence. We characterize the same superconducting qubit in two separate experimental configurations: in a printed-circuit board (PCB) package mounted within a standard cryogenic environment with no additional shielding, and then in nominally the same setup with the only change of embedding the entire PCB and qubit device in absorptive material. We find, surprisingly, that the embedding of highly sensitive quantum circuits in an absorptive, lossy medium can be useful for shielding and attenuating losses caused by radiation and  EM environmental factors within the cryogenic setup.\cite{Barends2011,note1} Our experimental procedure isolates the qubit loss mechanisms to be only those which are material related or on-chip/on-package, and linked to a bath that is thermalized to the base temperature.

The experimental device tested is a capacitively shunted flux qubit~\cite{Steffen10} (CSFQ) in the circuit quantum electrodynamics architecture. The qubit is capacitively coupled to a coplanar waveguide $\lambda/2$ resonator ($\omega_{\text{cav}} /2\pi= 10.3$ GHz), with a 6 $\mu$m center strip and 3 $\mu$m spacing to ground, via a $C_{\rm{qr}}\sim5$ fF interdigitated capacitor. A vacuum Rabi experiment (not shown) gives a qubit-resonator coupling strength $g/\pi\sim200$~MHz. The resonator is also capacitively coupled to a microwave feed line ($C_{\rm{c}}\sim2$~fF), with a linewidth $\kappa/2\pi = 470$~kHz, corresponding to $Q = 22,000$. 

\begin{figure}[tb!]
\includegraphics{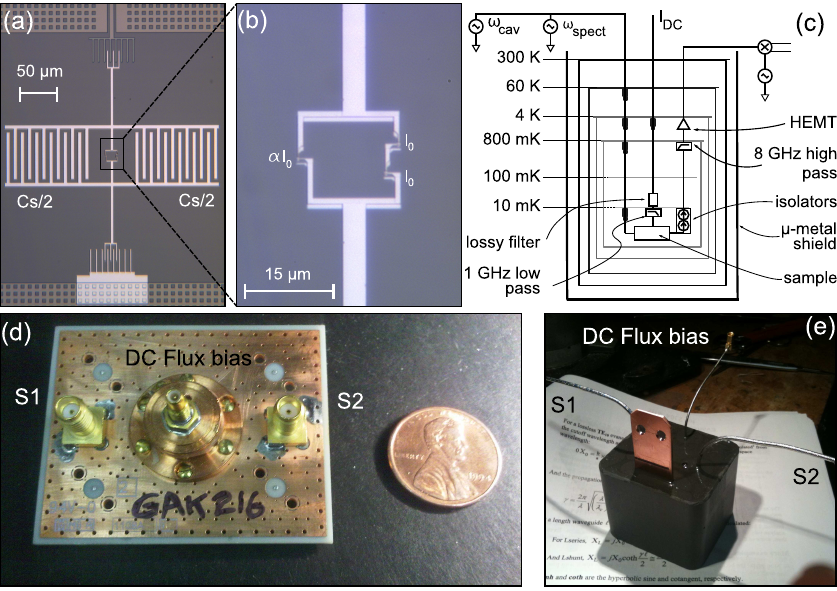}
\caption{(a)~Optical micrographs of the qubit device. (b)~Zoom in of qubit loop. From qubit spectroscopy, we fit to the qubit Hamiltonian and obtain the parameters $I_{0}=0.3$ $\mu$A, $\alpha=0.41$ and $C_{\rm{s}}=93$ fF. (c)~Schematic of experimental setup, showing relevant components on the three control lines to the qubit as well as the different temperature stages and shields.. (d)~PCB sample package for qubit experiments. Two rf lines, $S_1$ and $S_2$, connect to an on-chip feedline to address the cavity and qubit. A $\sim1000$-turn DC coil is mounted on the PCB for applying magnetic flux. (e)~Device after embedding in absorptive epoxy.}\label{fig1}
\end{figure}

The CSFQ consists of a 15 $\mu$m-wide square loop with three Josephson junctions, one with a smaller critical current $I_0$ than the other two by a ratio $\alpha$, and a shunt capacitor with 5 $\mu$m-wide fingers and gaps [Fig.~\ref{fig1}(a-b)]. More fabrication details of the qubit are described in previous work\cite{Steffen10,SteffenJPCM}. 

The sample is mounted in a custom designed multilayer copper clad FR-4 PCB, and flux biased with a hand-wound $\sim$1000-turn coil (Cu clad Nb/Ti wire) also attached to the PCB [Fig.~\ref{fig1}(d)]. In our first experiment, the sample package is attached to the mixing chamber of a cyrogen-free dilution refrigerator (DR) with a nominal base temperature of $\sim 15$ mK. Figure~\ref{fig1}(c) shows a schematic of the DR setup and all relevant experimental components anchored to different temperature stages. 

\begin{figure}[tb!]
\includegraphics[width=0.45\textwidth]{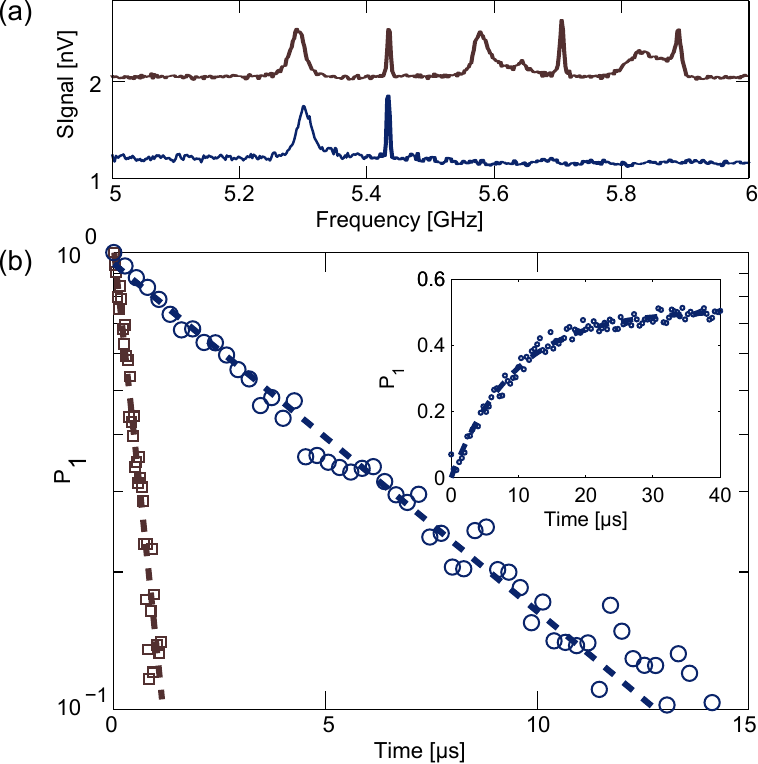}
\caption{\label{fig3}(a) Qubit spectroscopy at $0.51\Phi_0$ before (upper trace) and after (lower trace) embedding the sample in eccosorb. The three narrow peaks correspond to  multi-photon transitions. The broader peaks are the qubit bare transitions. The population of excited states in the qubit, arising from thermal excitations, is absent in the lower trace. (b) Energy relaxation time $T_1$ before (squares) and after (circles) embedding. Spin echo experiment (inset) gives $T_2=9.4\mu$s.}%
\end{figure}

In this configuration, spectroscopy of the qubit reveals significant steady-state thermal population. A high-power spectrum taken at a flux $\Phi=0.51\Phi_0$ is shown in Fig.~\ref{fig3}(a) (upper trace), in which we can identify equilibrium population in at least two excited states; we observe transitions at $\omega_{01}/2\pi = 5.3$ GHz, $\omega_{12}/2\pi = 5.58$ GHz, and $\omega_{23}/2\pi = 5.83$ GHz, where $\omega_{ij}$ is the transition frequency from state $|i\rangle$ to $|j\rangle$ in the CSFQ energy spectrum. The additional sharper peaks correspond to two-photon transitions between non-adjacent energy levels.

Time domain measurements are performed at the minimum transition frequency 5.01 GHz (``sweet spot'') located at half-integer flux-quanta. We measure an energy relaxation time $T_1 = 513$ ns for this bias point, obtained from a sliding $\pi$-pulse experiment (Fig.~\ref{fig3}(b), squares).
  
The spectroscopy data indicate that in the first experimenal configuration, our qubit is strongly coupled to some external energy. We performed different experiments to diagnose the source of this additional energy, first by adding and changing components in the experimental wiring in Fig.~\ref{fig1}(c). Additional attenuation to all the different stages, additional high- and low-pass filters on the input and output lines, improved heat sinking of all components, and different bandwidth isolators/circulators did not result in any noticeable qualitative improvement to heating or coherence times. However, we did observe a reduction in the effective heating on another qubit sample when it was mounted on a PCB and housed in a copper enclosure. That phenomenological result combined with recent work from the UCSB group\cite{BarendsAPS2011} suggesting increased resonator losses due to infrared radiation motivated us to completely eliminate the effect of external radiative and environmental energy sources. 

To protect the previously measured qubit and its sample package from external radiation, we place it in a mold and submerge it in CR-124 ECCOSORB epoxy (Emerson $\&$ Cuming). The resin is cured at 70$^\circ$C for over 12 hours. Flexible coaxial lines are used to provide connections to the input and output of the feedline as well as to the DC coil and a copper post is added for mechanical support and thermal anchoring [Fig.~\ref{fig1}(e)] prior to embedding.

Without changing any components interior to the cryostat setup shown in Fig.~\ref{fig1}(c), we recharacterize the sample and find substantial improvements to the qubit properties.  In fig.~\ref{fig3}(a), the qubit spectroscopy at the same $\Phi = 0.51\Phi_0$ flux bias shows remarkably fewer peaks. In the new experiment, only the $|0\rangle \rightarrow |1\rangle$ and two-photon $|0\rangle \rightarrow |2\rangle$ transitions are visible.

The qubit relaxation and coherence times also improve dramatically as a result of the embedding. 
$T_1^{\text{b}}$ (we use the superscript `b' to denote the results post embedding) of the qubit biased to the flux sweet spot increases by a factor of 10, from 513 ns previously to 5.7 $\mu$s [Fig.~\ref{fig3}(b)]. A Ramsey fringe experiment is performed giving a decoherence time $T_2^{*,\text{b}}= 5.6 \mu$s. However, by performing a spin echo experiment, we extract $T_2^{\text{b}} = 9.4$ $\mu$s, nearly twice $T_1^{\text{b}}$ (Fig.~\ref{fig3}(b) inset). These results imply a long pure dephasing time, consistent with recent experiments of Josephson junction qubits.\cite{Paik2011} The measurement that $T_2^{\text{b}} \neq T_2^{*,\text{b}}$ could be due to residual $1/f$ flux noise~\cite{Yoshihara2006}. 

Whereas 5.7 $\mu$s was the highest $T_1^{\text{b}}$ value measured, repeated measurements over a period of a week yielded values for $T_1^{\text{b}}$ between $\sim3.5$ and $\sim5.7$ $\mu$s, accompanied with similar fluctuations in $T_2^{*,\text{b}}$. We hypothesize that this variation in relaxation times could be due to slow fluctuations in dielectric loss. Future experiments will address this possibility.

The dramatic improvement in the qubit relaxation rate and the reduction of the effective qubit temperature give us clues about the mechanism and source of the external losses. Inside the DR, although the sample is firmly linked to the $\sim 15$ mK mixing chamber plate, the still shield surrounds the entire sample space, providing a source of 800 mK blackbody radiation to impinge upon the sample package.

We consider two physical processes by which radiation inside the DR can result in a thermally excited equilibrium qubit state and account for the observed relaxation time reduction. First, there is the mechanism of direct qubit excitation at 5 GHz due to blackbody radiation. The mean thermal photon number at 5 GHz due to an 800 mK radiator is estimated to be $\sim 3$. Although this could explain the additional transitions observed from the spectroscopy data, it does not account for the reduction in the relaxation time. To match our tenfold coherence time improvement, we estimate that the effective temperature of the bath~~\cite{schoelkopf02} would have to have been $\sim1.3$ K, higher than the still shield temperature. Furthermore, this would imply that the relaxation times are limited by direct radiation of the qubit at 5 GHz, which is inconsistent with simple calculations~\cite{Koch2007}.  

To account for the difference in relaxation times in the two experiments, we also consider the external loss process from the generation of quasiparticles by radiation of energy ($>80$ GHz) which exceeds the superconducting gap of Al ($\Delta \sim 200$ $\mu$eV). These quasiparticles can have a detrimental effect on the performance of the qubit through dissipation\cite{Lenander2011} and tunneling across the junctions.\cite{Catelani2011} 

\begin{figure}[tb!]
\includegraphics{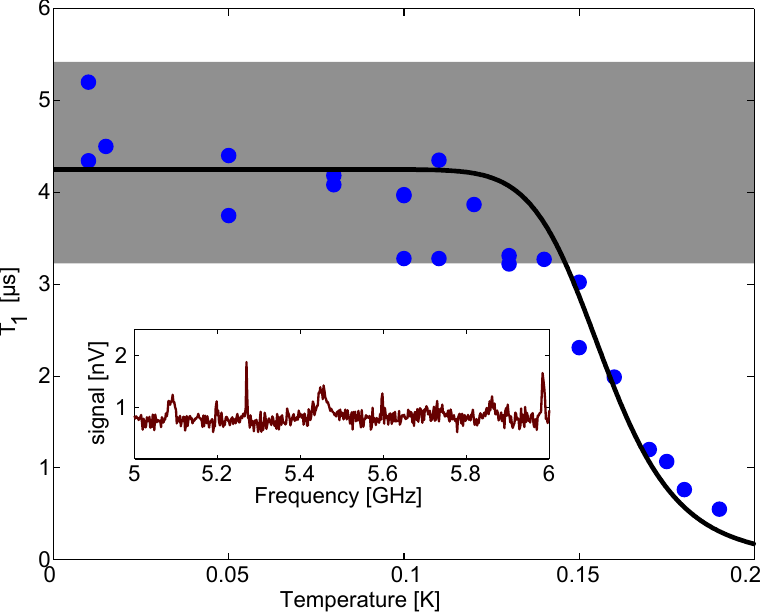}
\caption{$T_1$ versus mixing chamber temperature. Shaded area indicates range of $T_1$  consistent with repeated measurements. The drop off of $T_1$ above 150 mK is in agreement with quasiparticle generation (theory solid line). Spectroscopy (inset) and $T_1$ at 175 mK are in good qualitative agreement with the first experiment prior to the embedding process.}\label{fig4}
\end{figure} 

To mimic the effect of quasiparticles generated by radiation in the first experiment, we repeat spectroscopy and relaxation time measurements on the embedded sample at higher mixing chamber temperatures. Figure~\ref{fig4} shows $T_1$ versus sample temperature, with a roll-off around 140 mK. This behavior with increased temperature is qualitatively in agreement with recent theory~\cite{Catelani2011} of qubit relaxation due to quasiparticles (Fig.~\ref{fig4} solid line). At 175 mK, we find $T_1 \sim 700$ ns and a spectrum which shows up to 3 transitions (Fig.~\ref{fig4} inset), reminiscent of the original experiment. Therefore, prior to the embedding procedure, we attribute both the degraded lifetimes and the heating of the qubit to quasiparticles produced by radiation within the DR. 

Although the embedded qubit has a quality factor of $\sim 1.8\times 10^5$, this is still substantially below the limit placed by spontaneous decay through the resonator.\cite{Purcell,Houck2008} Candidates for the origin of the actual limit include spurious coupling to unwanted EM modes in the package\cite{Wenner2011} and dielectric losses arising from native oxides on the Al defining the qubit shunt capacitors.\cite{Steffen10,Paik2011} Future experiments will address these issues in greater detail.

To conclude, we have demonstrated that environmental radiation can have severe detrimental effects on a superconducting qubit, both in terms of coherence and effective temperature, most likely due to the generation of quasiparticles. With the simple procedure of embedding the sample package in a lossy epoxy material, the same qubit has remarkably improved parameters. This result is a major step for improving coherence and paves the way for determining intrinsic loss mechanisms in superconducting qubits.

We acknowledge engaging discussions with Blake Johnson and Chris Lirakis. The views expressed are those of the authors and do not reflect the official policy or position of the Department of Defense or the U.S. Government.

\end{document}